\documentclass[preprint,5p,times,twocolumn]{elsarticle}

\usepackage{dcolumn}
\usepackage{bm}
\usepackage{ifpdf}
\usepackage{hyperref}
\usepackage{xcolor,color,graphicx,graphics,physics}
\usepackage[spanish,english]{babel}
\usepackage[latin1]{inputenc}
\usepackage[OT1]{fontenc}
\usepackage{latexsym,amssymb,amsmath,amsfonts, slashed,cancel}
\usepackage{makeidx}
\usepackage{epsfig,subfigure}
\usepackage{natbib}
\usepackage{epstopdf}
\usepackage{mathrsfs}
\usepackage{hyperref}
\hypersetup{colorlinks=true, linkcolor=blue, citecolor=blue, urlcolor=blue}
\usepackage{enumerate}

\usepackage{fixmath}

\everymath{\displaystyle}
\usepackage{graphicx}

\usepackage[T1]{fontenc}
\usepackage{amsmath}
\usepackage{amssymb}
\usepackage{graphicx}
\usepackage{xcolor}

\newcommand{\bea}{\begin{eqnarray}}
\newcommand{\eea}{\end{eqnarray}}

\newcommand{\orcid}[1]{\href{https://orcid.org/#1}{\includegraphics[width=10pt]{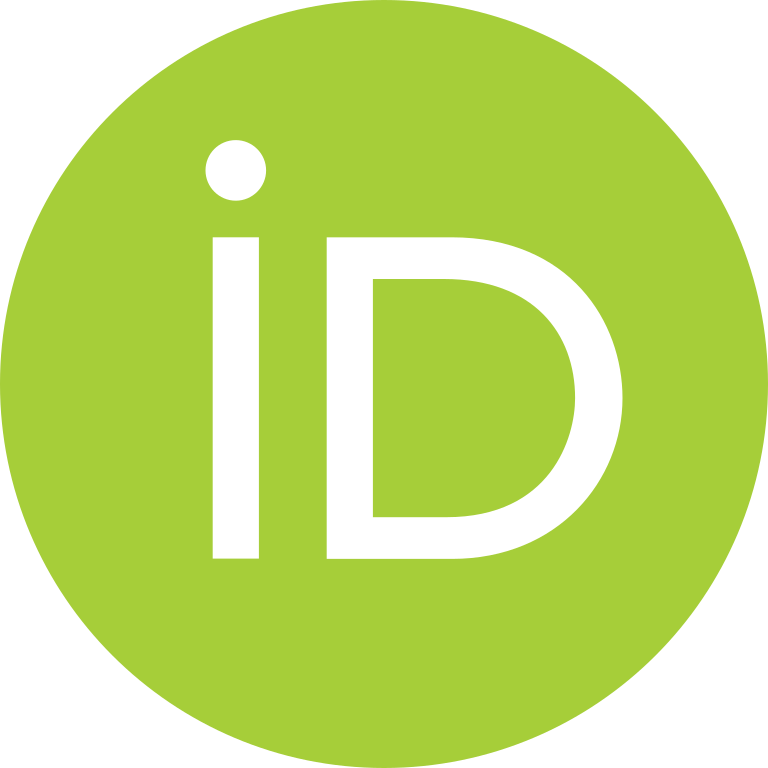}}}

\journal{Physics Letters B}

\begin{document}

\begin{frontmatter}

\title{Electron-positron scattering at finite temperature in Podolsky electrodynamics}

\author[UFMT]{D. S. Cabral  \orcid{0000-0002-7086-5582}}
\ead{danielcabral@fisica.ufmt.br}

\author[UFMT]{L. A. S. Evangelista \orcid{0009-0002-3136-2234}}
\ead{lucassouza@fisica.ufmt.br}

\author[UFMT]{L. H. A. R. Ferreira \orcid{0000-0002-4384-2545}}
\ead{luiz.ferreira@fisica.ufmt.br}

\author[UFMT]{A. F. Santos \orcid{0000-0002-2505-5273}}
\ead{alesandroferreira@fisica.ufmt.br}

\address[UFMT]{Programa de P\'{o}s-Gradua\c{c}\~{a}o em F\'{\i}sica, Instituto de F\'{\i}sica, 
Universidade Federal de Mato Grosso, Cuiab\'{a}, Brasil}

\begin{abstract}

The electron-positron annihilation process is investigated within the framework of Podolsky's generalized electrodynamics at finite temperature. In this theory, a higher-derivative term modifies the photonic kinetic sector, introducing a massive mode while preserving gauge invariance. Thermal effects are incorporated using the real-time Thermo Field Dynamics formalism. The total cross section is calculated, and the individual contributions of the Podolsky parameter and thermal effects are analyzed to highlight their influence on the scattering process.

\end{abstract}


\end{frontmatter}

\section{Introduction}

Maxwell's electrodynamics is the fundamental theory describing electromagnetic phenomena and serves as the basis for all scientific advancements in the field. Generalizations of Maxwell's theory exist, such as Proca electrodynamics, which introduces a mass term for the vector field. This modification breaks gauge invariance; nonetheless, Proca's theory successfully describes massive vector bosons like the $Z$ and $W^{\pm}$ bosons within the Standard Model (SM) \cite{qf,iqf}.

A natural generalization of Maxwell's theory is provided by Podolsky's generalized electrodynamics, which introduces a second-order derivative term coupled to the Podolsky parameter. This parameter generates a massive photon sector, similar to Proca's theory, but with the advantage of preserving gauge invariance. Proposed in 1942, Podolsky's theory was developed to regularize the high-frequency divergences present in Maxwell's electrodynamics \cite{p1}. For example, the electrostatic potential of a point charge diverges as $r \to 0$ in Maxwell's theory, an issue resolved within Podolsky's framework by yielding a finite and non-zero value at the origin \cite{p2,p3}. Furthermore, Frenkel employed Podolsky's theory to resolve the longstanding ``4/3 problem of classical electrodynamics,'' which remains unsolved in Maxwell's formulation \cite{43}.

Despite the similarity between the descriptions, the massive sector introduced by the Proca-like theory treats the photon mass as a small parameter appearing quadratically in the gauge field \cite{greiner}. Recent experimental constraints on the photon mass from this approach include upper limits of approximately $2.9 \times 10^{-14}$ eV from fast radio bursts dispersion analysis \cite{mass1}, $1.1 \times 10^{-12}$ eV from studies of frequency-dependent dispersion of intergalactic pulsars \cite{mass2}, and $5.34 \times 10^{-10}$ eV from deviations in the classical dispersion relation observed in fast radio bursts from cosmological sources \cite{mass3}.
In contrast, the massive sector arising from the Podolsky-like Lagrangian results from a correction in the kinetic term, characterized by a parameter $\lambda = 1/m$, which sets a lower bound on the photon mass. Several estimates for this lower bound have been proposed: approximately $4$ MeV based on comparisons between cosmic microwave background radiation data and the Stefan-Boltzmann law \cite{pod1}; about $35.51$ MeV derived from the hydrogen atom ground state, considering deviations from the Coulomb potential \cite{pod2}; and near $37$ GeV inferred from the expansion of the primordial universe within a generalized electrodynamics framework \cite{pod3}.

The SM of particle physics is the most successful framework currently available for describing the fundamental interactions among the basic constituents of matter, excluding gravity. It unifies the strong, weak, and electromagnetic forces through the exchange of gauge bosons that mediate these interactions. Quantum Electrodynamics (QED), a subset of the SM describing electromagnetic phenomena, is renowned for its exceptional predictive accuracy, extensively confirmed by experiments. One of the classical tests of QED involves scattering processes between charged particles, with electron-positron scattering being particularly significant due to its conceptual simplicity and phenomenological relevance \cite{qed}. This process, which proceeds via annihilation channels or virtual photon exchange, is accurately described by Feynman diagrams, enabling precise calculations of differential cross sections. Electron-positron scattering has been extensively studied across various sectors of the SM \cite{iqf, greiner, qed}, experimentally tested \cite{exp1, exp2, exp3, exp4}, and explored in numerous theoretical extensions \cite{lv1, lv2}. In this work, the $e^- + e^+ \to \ell^- + \ell^+$ scattering is investigated within the framework of Podolsky's generalized electrodynamics, where the photon propagator includes an additional term that necessitates its reformulation. Furthermore, thermal effects are incorporated to provide a more complete and realistic description of interactions, reflecting the intrinsic nature of physical systems in the universe. Although there are works that address real-time thermal electron-positron scattering \cite{our2, our3}, this paper is among the first to investigate QED scattering within Podolsky's theory at finite temperature, analyzed in the framework of the TFD formalism.

Several approaches have been developed to study quantum field theory under thermal effects. Among them, the Thermo Field Dynamics (TFD) formalism plays a prominent role in the analysis of tree-level scattering processes. This formalism allows quantum systems to be described using tools from statistical physics and has been extensively studied in the literature \cite{our1, our4, our5}. In TFD, vacuum expectation values are expressed as statistical averages of the system. The formalism is based on the construction of a thermal Hilbert space, $\mathcal{S}_\beta$, which is obtained by doubling the standard Hilbert space as $\mathcal{S}_\beta = \mathcal{S} \otimes \tilde{\mathcal{S}}$, where $\tilde{\mathcal{S}}$ denotes the duplicated (tilde) Hilbert space \cite{thermofield}. A crucial element of this approach is the Bogoliubov transformation, which is employed to construct thermal states and redefine the creation and annihilation operators for bosons and fermions. These operators are expressed as linear combinations of thermal operators acting on linearly independent thermal states. Consequently, the thermal propagator for any given theory can be derived in the presence of thermal effects, enabling the study of various physical processes \cite{tqft, ume}.

This paper is organized as follows. In Section \ref{s2}, Podolsky's electrodynamics is introduced, with the momentum-space propagator for an arbitrary gauge presented in Subsection \ref{p1}. In Subsection \ref{s2.1}, the $e^- + e^+ \to \ell^- + \ell^+$ scattering at zero temperature is calculated using the results obtained previously, and the modifications to the cross section caused by the Podolsky parameter are analyzed. Section \ref{s3} presents the main concepts of the TFD formalism. In Subsection \ref{p2}, the Podolsky photon propagator at finite temperature is derived, and in Subsection \ref{s3.1}, it is applied to the $e^- + e^+ \to \ell^- + \ell^+$ scattering process at finite temperature, demonstrating the combined effects of generalized electrodynamics and thermal contributions on the cross section. Finally, concluding remarks are provided in Section \ref{s4}.

\section{Podolsky Electrodynamics}\label{s2}

In this section, the fundamental aspects of Podolsky electrodynamics are presented. A massive representation for the photon is introduced while preserving gauge invariance. This is achieved by including a second-order derivative term in the Lagrangian. The photon propagator is derived within this framework, and, as an application, the electron-positron scattering at zero temperature is investigated.

\subsection{Photon Propagator in Podolsky Electrodynamics}\label{p1}

Since the Lagrangian of generalized electrodynamics includes an additional contribution in the kinetic sector, the structure of the photon propagator is consequently modified. In this context, the objective is to analyze how this modification affects the propagator.

The total Lagrangian of generalized electrodynamics, as given in Ref.~\cite{p2}, is 
\begin{eqnarray}
    \mathcal{L}&=&-\frac{1}{4}F_{\mu\nu}F^{\mu\nu}+\frac{\lambda^2}{2}\partial_\mu F^{\mu\nu}\partial_\alpha F^\alpha_\nu-\frac{1}{2\xi}\left(\partial_\mu A^\mu\right)^2\nonumber\\
    &+&\overline{\psi}\left(i\gamma^\mu\partial_\mu-m\right)\psi-e\overline{\psi}\gamma^\mu A_\mu\psi,\label{lagran}
\end{eqnarray}
where $\lambda$ and $\xi$ are the Podolsky and gauge parameters, respectively. This choice of gauge-fixing Lagrangian is not unique; other commonly used terms can be found, for instance, in Ref.~\cite{bufalo}. Considering only the free-field contribution of the photon, the following equation of motion is obtained in position space
\begin{equation}
    \left( \Box g_{\mu\nu} - \partial_\nu \partial_\mu + \lambda^2 g_{\mu\nu} \Box \Box - \lambda^2 \Box \partial_\mu \partial_\nu + \frac{1}{\xi} \partial_\mu \partial_\nu \right)A^\mu(x)=0.
\end{equation}
The electromagnetic field is expanded as
\begin{align}
    A_\mu(x) = \int \frac{d^3k}{(2\pi)^3} N^\prime_k \sum_s \epsilon_\mu(k,s) \left[ a_{k,s} e^{-ik \cdot x} + a^\dagger_{k,s} e^{ik \cdot x} \right],\label{eq06}
\end{align}
with $\epsilon_\mu(k,\lambda)$ denoting the polarization vector, $k$ representing the photon momentum, $N^\prime_k=[2|\vec{k}|(1-\lambda^2k^2)]^{-1}$ the normalization constant for photons, and $a$ and $a^\dagger$ the annihilation and creation operators for the photon field. 

Therefore, in momentum space, the expression takes the form
\begin{equation}
    \left(-k^2g_{\mu\nu}+k_\mu k_\nu+\lambda^2 g_{\mu\nu}k^4-\lambda^2 k^2 k_\mu k_\nu-\frac{1}{\xi}k_\mu k_\nu\right)\epsilon^\mu(k)=0,\label{eq01}
\end{equation}
which leads to the following expression for the photon propagator in generalized electrodynamics
\begin{equation}
    D_F^{\mu\nu}(k^2)=-\frac{i}{k^2(1-\lambda^2k^2)}\left[g_{\mu\nu}-(1-\xi+\xi \lambda^2k^2)\frac{k_\mu k_\nu }{k^2}\right],\label{eq03}
\end{equation}
as derived in Ref.~\cite{malufprop}.  According to the dispersion relation, the propagator exhibits two poles: one associated with standard QED and another arising from Podolsky's generalization. The first is not problematic due to the off-shell nature of virtual particles, while the second emerges naturally from the theory.

The following completeness relation is defined here
\begin{equation}
    \sum_s\epsilon_s^{\mu}(k)\epsilon^{\nu*}_s(k)=g^{\mu\nu}+\lambda^2{k}^{\mu}{k}^{\nu}.\label{eq08}
\end{equation}
An analogous relation has been derived in the context of massive photons in Maxwell-Proca theory, as reported in Ref.~\cite{greiner}.

It should be noted that, by taking the limit $\lambda \to 0$, the standard photon propagator from QED is recovered. Furthermore, although the formulation is developed in an arbitrary gauge, without loss of generality, the Feynman gauge $\xi = 1$ is adopted for the remainder of this paper.  Since the theory is gauge invariant by construction, any value of $\xi$ can be chosen; in other words, the currents remain conserved. in  With the modified propagator thus obtained, the $e^- + e^+ \to \ell^- + \ell^+$ scattering process is described, and its effect on the interaction cross section is analyzed.

\subsection{The $e^- + e^+ \to \ell^- + \ell^+$ scattering process at zero temperature}\label{s2.1}

\begin{figure}[!h]
    \centering
    \includegraphics[width=0.54
    \linewidth]{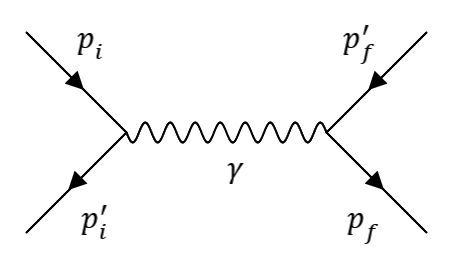}
    \caption{Feynman diagram for the $e^- + e^+ \to \ell^- + \ell^+$ scattering process. Here, $p_i$ ($p_i^\prime$) and $p_f$ ($p_f^\prime$) denote the initial and final momenta of the particle (antiparticle), respectively.
}
    \label{feynmandiag}
\end{figure}

This section is dedicated to the study of the scattering process $e^- + e^+ \to \ell^- + \ell^+$ in the absence of thermal effects. The corresponding Feynman diagram is illustrated in Figure \ref{feynmandiag}.

The transition amplitude is defined as
\begin{align}
\mathcal{M} = \bra{f} S^{(2)} \ket{i}, \label{M}
\end{align}
which encodes the probability amplitude for the interaction to occur, depending on the asymptotic initial and final states, as well as on the scattering matrix $S^{(n)}$. Since the calculation is performed at tree level, $n=2$ is set. The initial and final states are expressed as
\begin{align}
    \ket{i} &= \ket{e^-_i, e^+_i} = b^{\dagger}_p c^\dagger_k \ket{0}, \\
    \ket{f} &= \ket{\ell^-_f, \ell^+_f} = b^{\dagger}_{p^\prime} c^\dagger_{k^\prime} \ket{0},
\end{align}
where $p$ ($k$) and $p^\prime$ ($k^\prime$) denote the initial (final) momenta of the particle (antiparticle), respectively.

The second-order scattering matrix is defined as
\begin{align}
    S^{(2)} = -\frac{1}{2} \int\int d^4x d^4y \; \mathcal{T}[\mathcal{L}_I(x)\mathcal{L}_I(y)],
\end{align}
where $\mathcal{T}$ denotes the time-ordering operator, and $\mathcal{L}_I$ represents the interaction Lagrangian. In the present context, it takes the form\begin{align}
    \mathcal{L}_I = -e \overline{\psi} \gamma^\mu A_\mu \psi,
\end{align}
with $e$ being the electron charge, $\psi$ the Dirac spinor field, $\overline{\psi} = \psi^\dagger \gamma^0$, and $A_\mu$ the electromagnetic four-potential.

By substituting these expressions into the transition amplitude \eqref{M}, the following is obtained
{\small
\begin{align}
    \mathcal{M} = -\frac{e^2}{2} \int\int d^4x  d^4y \bra{f} \mathcal{T}[\overline{\psi}(x) \gamma^\mu A_\mu(x) \psi(x) \overline{\psi}(y) \gamma^\nu A_\nu(y) \psi(y)] \ket{i}.\label{M1}
\end{align}}

The fermionic field operators are expressed in terms of plane-wave solutions as
\begin{align}
    \psi(x) &= \int \frac{d^3p}{(2\pi)^3} N_p \sum_s \left[ u(p,s) b_{p,s} e^{-ip \cdot x} + v(p,s) c^\dagger_{p,s} e^{ip \cdot x} \right], \\
    \overline{\psi}(x) &= \int \frac{d^3p}{(2\pi)^3} N_p \sum_s \left[ \overline{u}(p,s) b^\dagger_{p,s} e^{ip \cdot x} + \overline{v}(p,s) c_{p,s} e^{-ip \cdot x} \right],
\end{align}
where $N_p$ is the normalization constant, $u$ and $v$ denote the Dirac spinors, and $b$, $b^\dagger$, $c$, and $c^\dagger$ represent the annihilation and creation operators for particles and antiparticles, respectively.

To analyze the $e^- + e^+ \to \ell^- + \ell^+$ scattering process, only the field components contributing to the Feynman diagram shown in Figure \ref{feynmandiag} are considered. In other words, by substituting the fermionic fields into Eq.~\eqref{M1} and selecting the spinor terms relevant to the scattering, the transition amplitude can be expressed as
\begin{eqnarray}
    \mathcal{M}&=&-\frac{e^2}{2}\int\int d^4x\,d^4y\;\sum_s e^{-ix(p_i+p^\prime_i)}e^{iy(p_f+p^\prime_f)}\nonumber\\
    &\times&\left[\overline{v}(p^\prime_i)\gamma^\mu u(p_i)\overline{u}(p_f)\gamma^\nu v(p^\prime_f)\right]
    \bra{0}A_\mu(x)A_\nu(y)\ket{0},
\end{eqnarray}
where the following relation has been used
\begin{eqnarray}
    &&\bra{0}b_{p_f} c_{p^\prime_f}\mathcal{T}[c_{p_1}b_{p_2}b^\dagger_{p_3}c^\dagger_{p_4}]b^\dagger_{p_i}c^\dagger_{p^\prime_i}\ket{0}\\
    &&=\frac{1}{N_{p_n}^4}\delta^3(p_f-p_3)\delta^3(p^\prime_f-p_4)\delta^3(p_1-p^\prime_i)\delta^3(p_2-p_i).\nonumber
\end{eqnarray}
It is important to note that the vacuum expectation value of the photonic field corresponds to the photon propagator in the Podolsky formulation of electromagnetism, given by Eq.~\eqref{eq03}. Using the identity for the Dirac delta function,
{\small
\begin{align}
    \int d^4x\,d^4y\;e^{ix(p_i+p^\prime_i-q)}e^{-iy(p_f+p^\prime_f-q)}=\delta^4(p_i+p^\prime_i-q)\delta^4(p_f+p^\prime_f-q),
\end{align}}
the transition amplitude can be written as
\begin{align}
    \mathcal{M}=\frac{ie^2(g^{\mu\nu}-\lambda^2 q^\mu q^\nu)}{2q^2(\lambda^2 q^2-1)}\left[\overline{v}(p^\prime_i)\gamma^\mu u(p_i)\overline{u}(p_f)\gamma^\nu v(p^\prime_f)\right].\label{M21}
\end{align}
This expression can be used to calculate the probability amplitude for the scattering. 

To calculate the cross section, the center-of-mass (CM) reference frame will be considered, where the momenta are given by
\begin{align}
    p^\prime_i&=(E,0,0,-E),\\
    p_i&=(E,0,0,E),\\
    p^\prime_f&=(E,-E\sin{\theta},0,-E\cos{\theta}),\\
    p_f&=(E,E\sin{\theta},0,E\cos{\theta}),
\end{align}
where $E$ is the energy and $\theta$ represents the scattering angle.  In this reference frame, the Mandelstam variables are given by
\begin{align}
    s=(2 E)^2;\quad t=-2 E^2(1-\cos\theta);\quad u=-2E^2(1+\cos\theta).
\end{align}
Furthermore, by definition, the differential cross section in the ultrarelativistic limit, i.e., when $\{m, m_\ell\} \to 0$, is given by
\begin{align}
    \left(\frac{d\sigma}{d\Omega}\right)=\frac{1}{64 \pi^2 s}\langle|\mathcal{M}|^2\rangle,\label{diffsigma}
\end{align}
with $\langle|\mathcal{M}|^2\rangle$ being the probability amplitude, which can be written as
\begin{align}
    \langle|\mathcal{M}|^2\rangle=\frac{1}{4}\sum_s |\mathcal{M}|^2
\end{align}
with the sum over $s$ running through the unpolarized helicity states.

\begin{figure}[ht]
    \centering
    \includegraphics[width=0.8\linewidth]{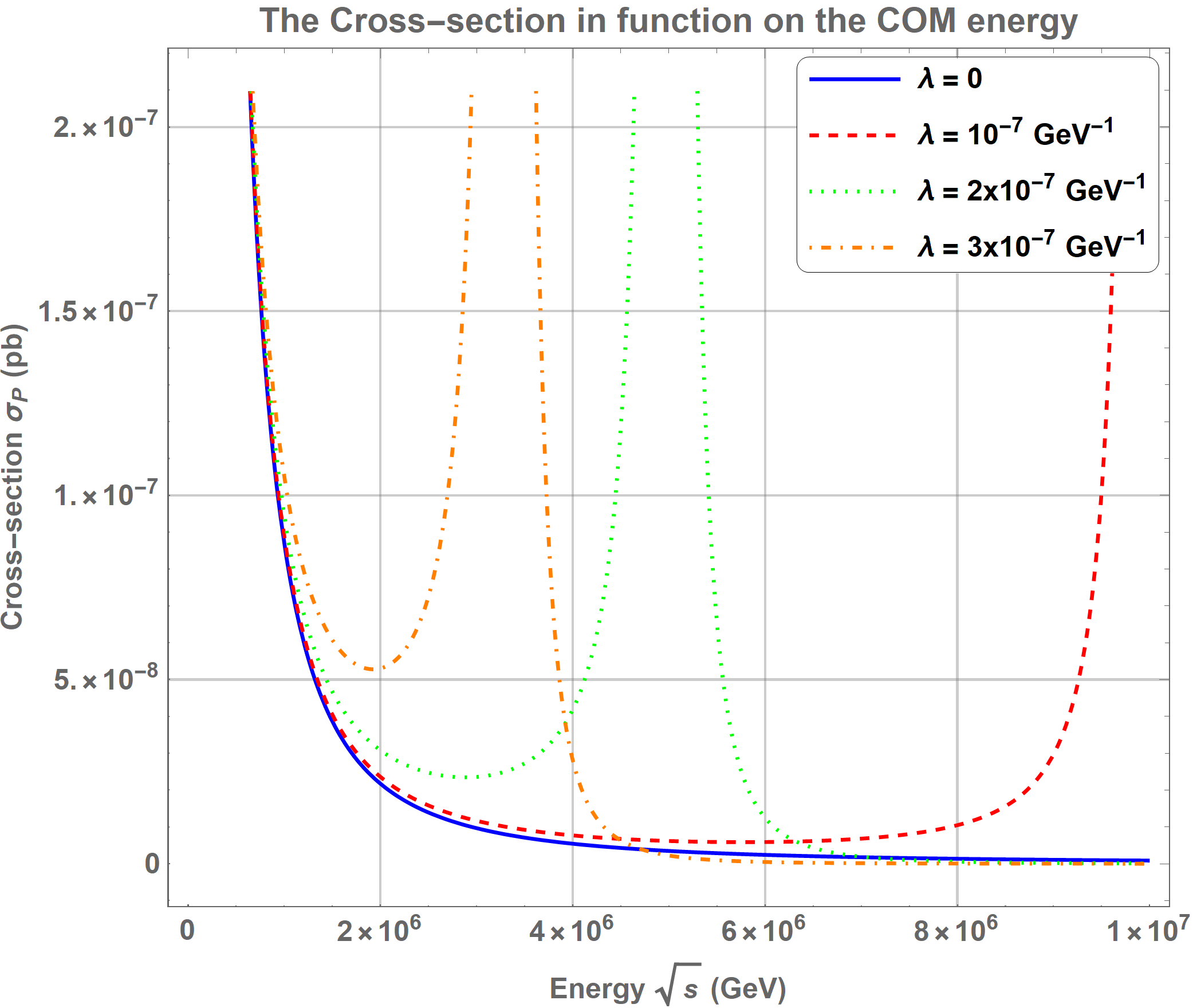}
    \caption{Scattering cross section given by Eq.~\eqref{sigma} as a function of the center-of-mass energy $\sqrt{s}$, shown for $\lambda = 0$ (blue solid line), $\lambda = 10^{-7}\,\text{GeV}^{-1}$ (red dashed line), $\lambda = 2 \times 10^{-7}\,\text{GeV}^{-1}$ (green dotted line), and $\lambda = 3 \times 10^{-7}\,\text{GeV}^{-1}$ (orange dot-dashed line). }
    \label{fig1}
\end{figure}

By inserting the transition amplitude from Eq.~\eqref{M21} into the probability amplitude, the following expression is obtained
{\small
\begin{align}
     \langle|\mathcal{M}|^2\rangle&=\frac{e^4}{4s^2(\lambda^2 s-1)^2}(g^{\mu\nu}-\lambda^2 q^\mu q^\nu)(g^{\alpha\xi}-\lambda^2 q^\alpha q^\xi)\nonumber\\
     &\times\sum_s\left[\overline{v}(p^\prime_i)\gamma^\mu u(p_i)\overline{u}(p_f)\gamma^\nu v(p^\prime_f)\right][\overline{v}(p^\prime_f)\gamma^\alpha u(p_f)\overline{u}(p_i)\gamma^\xi v(p^\prime_i)],
\end{align}}
where $s = (2E)^2$ represents the center-of-mass energy of the interaction. This expression enables the application of the fermionic completeness relation,
\begin{align}
    \sum_s u(p,s)\overline{u}(p,s)=\slashed{p}+m, \label{completeness}
\end{align}
along with the identity
{\small
\begin{eqnarray}
   &&[\overline{v}(p^\prime_i)\gamma^\mu u(p_i)\overline{u}(p_f)\gamma^\nu v(p^\prime_f)][\overline{v}(p^\prime_f)\gamma^\alpha u(p_f)\overline{u}(p_i)\gamma^\xi v(p^\prime_i)]\\
    &&=\mathrm{Tr}[v(p^\prime_i)\overline{v}(p^\prime_i)\gamma^\mu u(p_i)\overline{u}(p_i)\gamma^\xi]\mathrm{Tr}[u(p_f)\overline{u}(p_f)\gamma^\nu v(p^\prime_f)\overline{v}(p^\prime_f)\gamma^\alpha],\nonumber\label{completeness2}
\end{eqnarray}}
which leads to the following result
\begin{align}
     \langle|\mathcal{M}|^2\rangle=\frac{e^4(\cos^2\theta+1)}{(1-\lambda^2s)^2}.\label{M2}
\end{align}
This result can be substituted into the expression for the differential cross section given in Eq.~\eqref{diffsigma}, yielding
\begin{align}
    \left(\frac{d\sigma}{d\Omega}\right)_P=\frac{e^4(\cos2\theta+3)}{512\pi^2 E^2(1-\lambda^2s)^2},
\end{align}
which can be integrated over the solid angle to obtain the total cross section of the scattering process
\begin{align}
    \sigma_P=\frac{4\alpha^2\pi}{3s(1-\lambda^2s)^2},\label{sigma}
\end{align}
where the fine-structure constant $\alpha$ has been introduced, providing a more conventional representation commonly used in the literature. Accordingly, Eq.\eqref{sigma} describes the $e^- + e^+ \to \ell^- + \ell^+$ scattering process at zero temperature within the framework of Podolsky electrodynamics. The total cross section given in Eq.\eqref{sigma} is plotted and compared in Figure~\ref{fig1} for different values of the Podolsky parameter.

As expected, in the limit $\lambda \to 0$, the standard QED scattering result is recovered \cite{iqf}. When the Podolsky parameter is taken into account, it introduces a pole in the cross section at $s = \lambda^{-2}$, enhancing the probability and, consequently, the effective interaction area as the center-of-mass energy approaches this value. This pole shifts to higher energies as $\lambda$ decreases. Furthermore, at energies beyond the pole, the cross section decreases more rapidly for larger values of $\lambda$, due to the presence of the factor $(1 - \lambda^2 s)^{-2}$.

It is important to note that the expression obtained for the s-channel in the present work coincides with that derived in Ref.\cite{bufalo} for a different reaction, namely Bhabha scattering. Furthermore, in that study, analyzed in a non-thermal framework with a different gauge-fixing contribution, the resulting Lagrangian differs from Eq.\eqref{lagran} and, consequently, different expressions for the propagator and completeness relation than those given by Eqs.~\eqref{eq03} and \eqref{eq08}, respectively.

 As mentioned in the introduction, there are several lower bounds on the photon mass, including comparisons with experimental CMB data \cite{pod1} and studies on deviations of the ground state proposed by \cite{pod2}. A more precise constraint is provided by \cite{pod3}, which finds an upper limit near $37$ GeV by analyzing features of the primordial universe expansion within Podolsky electrodynamics. This result implies that the pole of the cross-section occurs at $\sqrt{s} \sim 37$ GeV, placing the pole of $\sigma_P$ within the measurable range \cite{exp01, exp02, exp03} and, in principle, allowing its detection. Since no such evidence has been observed experimentally, this indicates that current limits on the photon mass are still insufficient to reveal this effect. Therefore, our choice of parameter values represents, in our view, the most appropriate estimate within this scenario. For the construction of the graphs that follow, we adopt the limit $\lambda = 3 \times 10^{-7}$ GeV$^{-1}$.

In the next section, the fundamental aspects of TFD will be addressed, with a focus on how the theory modifies the propagator in generalized electrodynamics. These concepts will subsequently be applied to calculate the scattering process at finite temperature.

\section{Thermo Field Dynamics formalism}\label{s3}

In this section, the fundamental concepts related to TFD will be presented, along with their application to the calculation of the $e^- + e^+ \to \ell^- + \ell^+$ scattering process at finite temperature.

\subsection{Theoretical concepts}\label{p2}

Among the various approaches available to incorporate thermal effects into field theory, the real-time formalism of Thermo Field Dynamics (TFD) stands out as particularly effective \cite{tqft}. In this framework, temperature is introduced by doubling the Hilbert space, that is, by defining the thermal space as $\mathcal{S}_\beta = \mathcal{S} \otimes \tilde{\mathcal{S}}$, where $\mathcal{S}$ and $\tilde{\mathcal{S}}$ represent the original Hilbert space and its copy (the tilde space), respectively. By employing specific linear combinations known as Bogoliubov transformations, quantum averages at finite temperature can be constructed. These concepts are fundamental for understanding how thermal effects influence physical systems, an essential consideration, given that our universe does not exist at zero temperature.

In practice, the thermal vacuum state is defined as
\begin{equation}
\ket{0(\beta)} =\frac{1}{\sqrt{Z(\beta)}}\sum_n\exp{-\frac{1}{2}\beta E_n}\ket{n,\tilde{n}},  
\end{equation}
where $Z(\beta)$ is the partition function and $\beta$ is the inverse temperature. Based on this definition, any thermal observable $O$ can be obtained as
\begin{equation}
    \braket{O}=\bra{0(\beta)}O\ket{0(\beta)}.
\end{equation}
For example, the thermal photonic propagator $D_F^{\mu\nu}(x - y; \beta)$ in generalized QED is defined as
\begin{align}
    D_F^{\mu\nu}(x-y;\beta)&=\bra{0(\beta)}\mathcal{T}\left[A^{\mu}(x)A^{\nu}(y)\right]\ket{0(\beta)}\label{eq05}
\end{align}
with $\mathcal{T}$ being the time-ordering operator. 

To determine the structure of the Green function in Eq.\eqref{eq05}, the bosonic field must be decomposed into plane waves, as shown in Eq.\eqref{eq06}. Since the creation and annihilation operators act in the space $\mathcal{S}$, while the expectation value is evaluated in the thermal space $\mathcal{S}_\beta$, the application of Bogoliubov transformations is required. These transformations are defined, for bosonic operators, as
\begin{align}
    a_{k,\lambda}&=U^\prime(k,\beta)a_{k,\lambda}(\beta)+V^\prime(k,\beta)\tilde{a}^\dagger_{k,\lambda}(\beta),\nonumber\\ a^\dagger_{k,\lambda}&=U^\prime(k,\beta)a^\dagger_{k,\lambda}(\beta)+V^\prime(k,\beta)\tilde{a}_{k,\lambda}(\beta),
\end{align}
and, for fermionic operators, as
\begin{align}
    b_{k,\lambda}&=U(k,\beta)b_{k,\lambda}(\beta)+V(k,\beta)\tilde{b}^\dagger_{k,\lambda}(\beta),\nonumber\\ b^\dagger_{k,\lambda}&=U(k,\beta)b^\dagger_{k,\lambda}(\beta)+V(k,\beta)\tilde{b}_{k,\lambda}(\beta).\label{bogoferm}
\end{align}
The coefficients are defined such that
\begin{eqnarray}
    V^{\prime 2}(k,\beta)&=&n(k,\beta);\quad\quad  U^{\prime 2}-V^{\prime 2}=1;\nonumber\\
     V^2(k,\beta)&=&f(k,\beta);\quad\quad U^2+V^2=1,\label{eq09}
\end{eqnarray}
where $n(k, \beta)$ and $f(k, \beta)$ are the Bose-Einstein and Fermi-Dirac distributions, respectively. The physical interpretation of the thermal field operators is obtained by considering the usual operators as immersed in a heat bath, so that they follow the corresponding thermal statistics.

With this in mind, the photon propagator at finite temperature can be written as
{\small
\begin{align}
    \bra{0(\beta)}A^{\mu}(x)A^{\nu}(y)\ket{0(\beta)}&=\int\frac{d^3k}{(2\pi)^3}N_k\sum_s\left[U^{\prime2}(k,\beta)\epsilon^\mu(k,s)\epsilon^{\nu*}(k,s)e^{-ik(x-y)}\right.\nonumber\\&\left.+V^{\prime2}(k,\beta)\epsilon^{\mu*}(k,s)\epsilon^{\nu}(k,s)e^{ik(x-y)}\right],\label{eq07}
\end{align}}
It is necessary to refer back to Eq.\eqref{eq08} to express the sum over polarizations. By substituting Eqs.\eqref{eq07} and \eqref{eq08} into Eq.~\eqref{eq05}, the following expression is obtained
\begin{eqnarray}
    D^{\mu\nu}_F(x-y;\beta)&=&\int\frac{d^3k}{(2\pi)^3}\frac{\Theta(x^0-y^0)}{2|\vec{k}|(\lambda^2k^2-1)}\nonumber\\
    &\times&\left[U^{\prime^2}e^{-ik(x-y)}+V^{\prime^2}e^{ik(x-y)}\right]\left[g^{\mu\nu}+\lambda^2 k^\mu k^\nu\right]\nonumber\\&+&\int\frac{d^3k}{(2\pi)^3}\frac{\Theta(y^0-x^0)}{2|\vec{k}|(\lambda^2k^2-1)}\nonumber\\
    &\times&\left[U^{\prime^2}e^{-ik(y-x)}+V^{\prime^2}e^{ik(y-x)}\right]\left[g^{\mu\nu}+\lambda^2k^\mu k^\nu\right].\nonumber\\
\end{eqnarray}
In this manner, by employing the relations in Eq.~\eqref{eq09} and performing the complex integration using the residue theorem, the following result is obtained
\begin{align}
    D_F^{\mu\nu}(x-y;\beta)&=\int\frac{d^4k}{(2\pi)^4}D_F(k;\beta)e^{-ik(x-y)}\label{propb}
\end{align}
with
\begin{align}
    D_F(k;\beta)=-i\left[\frac{1}{k^2(1-\lambda^2k^2)}+\frac{2\pi i n(k,\beta)\delta(k^2)}{(1-\lambda^2k^2)}\right]\left(g^{\mu\nu}+\lambda^2k^{\mu}k^{\nu}\right).
\end{align}
The preceding equation represents the thermal propagator for generalized electrodynamics. This propagator will be employed in the calculation of the probability amplitude for the $e^- + e^+ \to \ell^- + \ell^+$ scattering process at finite temperature. In the low-temperature limit, i.e., as $\beta \to \infty$ and $T \to 0$, the standard Podolsky propagator given by Eq.~\eqref{eq03} is recovered.

In the next section, the scattering process will be analyzed through the calculation of its probability amplitude and cross section.

\subsection{The $e^- +e^+ \to \ell^-+\ell^+$ scattering at finite temperature}\label{s3.1}

Here, the thermal scattering process is treated analogously to the zero-temperature case discussed previously. This is possible due to a key advantage of the TFD formalism: it allows thermal effects to be incorporated in a way that preserves the structure of zero-temperature quantum field theory. To this end, the analysis begins with the definition of the transition amplitude, given by
\begin{align}
    \mathcal{M}(\beta) = \bra{f(\beta)}\hat{S}^{(2)}\ket{i(\beta)}.\label{MB}
\end{align}
It is important to note that the asymptotic states now depend explicitly on temperature and are expressed as
\begin{align}
    \ket{i(\beta)} &= \ket{e^-_i, e^+_i}_\beta = b^{\dagger}_p(\beta) c^\dagger_k(\beta) \ket{0(\beta)}, \\
    \ket{f(\beta)} &= \ket{\ell^-_f, \ell^+_f}_\beta = b^{\dagger}_{p^\prime} (\beta)c^\dagger_{k^\prime} (\beta)\ket{0(\beta)}.
\end{align}
Furthermore, the scattering matrix now reflects the doubling of the Hilbert space inherent to the TFD formalism, and at tree level it is given by
\begin{align}
    S^{(2)} = -\frac{1}{2} \int\int d^4x d^4y \; \mathcal{T}[\hat{\mathcal{L}}_I(x)\hat{\mathcal{L}}_I(y)],
\end{align}
where the interaction Lagrangian is defined as
\begin{align}
    \hat{\mathcal{L}}_I=\mathcal{L}_I-\tilde{\mathcal{L}}_I=-e \overline{\psi} \gamma^\mu A_\mu \psi+e \tilde{\overline{\psi}} \gamma^{\mu*} \tilde{A}_\mu \tilde{\psi}.
\end{align}
It should be emphasized that the substitution of these quantities into the transition amplitude in Eq.~\eqref{MB} leads to contributions from the non-tilde fields, the tilde fields, and their mixed terms. However, only the non-tilde sector contributes to the physical observables of interest. Although analogous expressions arise from the tilde sector, they remain confined to the $\tilde{\mathcal{S}}$-space and, as such, do not correspond to any measurable physical quantities.

By substituting the relevant quantities into the transition amplitude and employing the thermal propagator of Podolsky theory given in Eq.~\eqref{propb}, the following expression is obtained
{\small
\begin{align}
    \mathcal{M}(\beta)=&\frac{ie^2 U^4(\beta)}{2}\left[\frac{1}{q^2(1-\lambda^2q^2)}+\frac{2\pi i n(q,\beta)\delta(q^2)}{(1-\lambda^2q^2)}\right]\left(g^{\mu\nu}+\lambda^2q^{\mu}q^{\nu}\right)\nonumber\\
    &\times\left[\overline{v}(p^\prime_i)\gamma^\mu u(p_i)\overline{u}(p_f)\gamma^\nu v(p^\prime_f)\right].
\end{align}}
In this result, the Bogoliubov transformation for the field operators, given in Eq.\eqref{bogoferm}, has been applied, introducing the factor $U$. In the zero-temperature limit, $\beta \to \infty$ (i.e., $T \to 0$), one finds that $n(q^2; \beta) \to 0$ and $U(\beta) \to 1$, so the zero-temperature transition amplitude in Eq.\eqref{M2} is recovered.

Thus, the probability density in the TFD formalism is expressed as
\begin{eqnarray}
    && \langle|\mathcal{M}(\beta)|^2\rangle=\frac{1}{4}\sum_s |\mathcal{M}(\beta)|^2\nonumber\\
     &=&\frac{e^4 U^8(\beta)}{16}\left[\frac{1}{s^2(1-\lambda^2s)^2}+\frac{4\pi^2  n^2(q,\beta)\delta^2(q^2)}{(1-\lambda^2s)^2}\right]\nonumber\\
     &\times&(g^{\mu\nu}-\lambda^2 q^\mu q^\nu)(g^{\alpha\xi}-\lambda^2 q^\alpha q^\xi)\nonumber\\
     &\times&\sum_s\left[\overline{v}(p^\prime_i)\gamma^\mu u(p_i)\overline{u}(p_f)\gamma^\nu v(p^\prime_f)\right]\nonumber\\
     &\times&\left[\overline{v}(p^\prime_f)\gamma^\alpha u(p_f)\overline{u}(p_i)\gamma^\xi v(p^\prime_i)\right].
\end{eqnarray}
By applying the relations in Eqs.~\eqref{completeness} and \eqref{completeness2}, the preceding equation simplifies to
{\small
\begin{align}
     \langle|\mathcal{M}(\beta)|^2\rangle=
     \frac{e^4(\cos^2\theta+1)}{(1-4\lambda^2 E^2)^2}\left(\frac{1+\tanh{\frac{\beta E}{2}}}{2}\right)^4\left(64 \pi^2 \delta^2(s) n^2(q;\beta) E^4+1\right),\label{MB1}
\end{align}}
which correctly reduces to the zero-temperature expression given in Eq.\eqref{M2}. For convenience in the thermal analysis, the factor
\begin{align}
    U^8(\beta)=\left(\frac{1+\tanh{\frac{\beta E}{2}}}{2}\right)^4,\label{eq11}
\end{align}
has been introduced.

\begin{figure}[ht]
    \centering
    \includegraphics[width=0.75\linewidth]{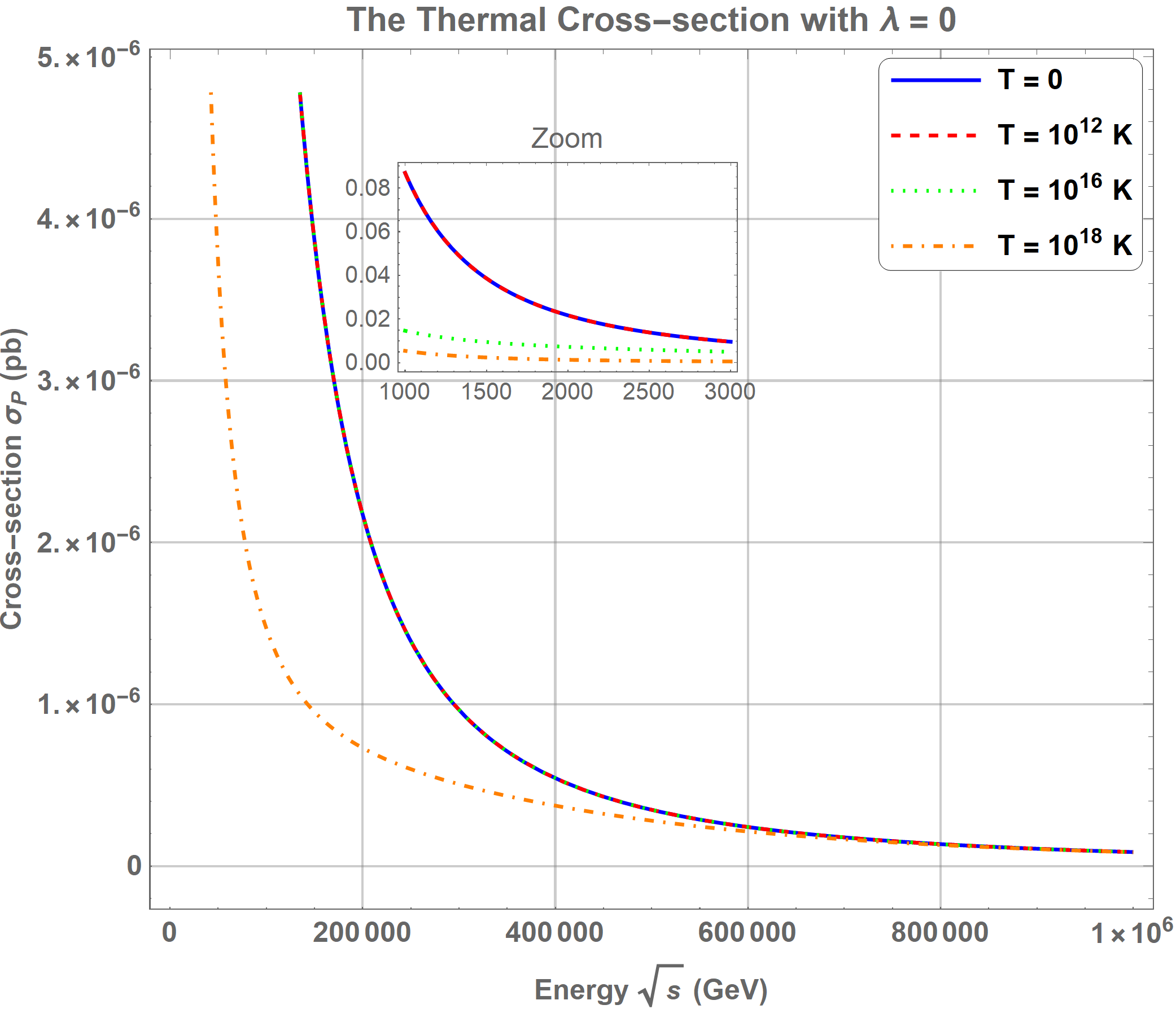}
    \caption{Thermal cross-section \eqref{eq10} versus center-of-mass energy $\sqrt{s}$ at $T=0$ (blue solid line), $10^{12}$ K (red dashed line), $10^{16}$ K (green dotted line), and $10^{18}$ K (orange dot-dashed line), with $\lambda=0$. The inset zooms in on the range $10^3~\mathrm{GeV} \leq \sqrt{s} \leq 3 \times 10^3~\mathrm{GeV}$.}
    \label{fig2}
\end{figure}

\begin{figure}[ht]
    \centering
    \includegraphics[width=0.75\linewidth]{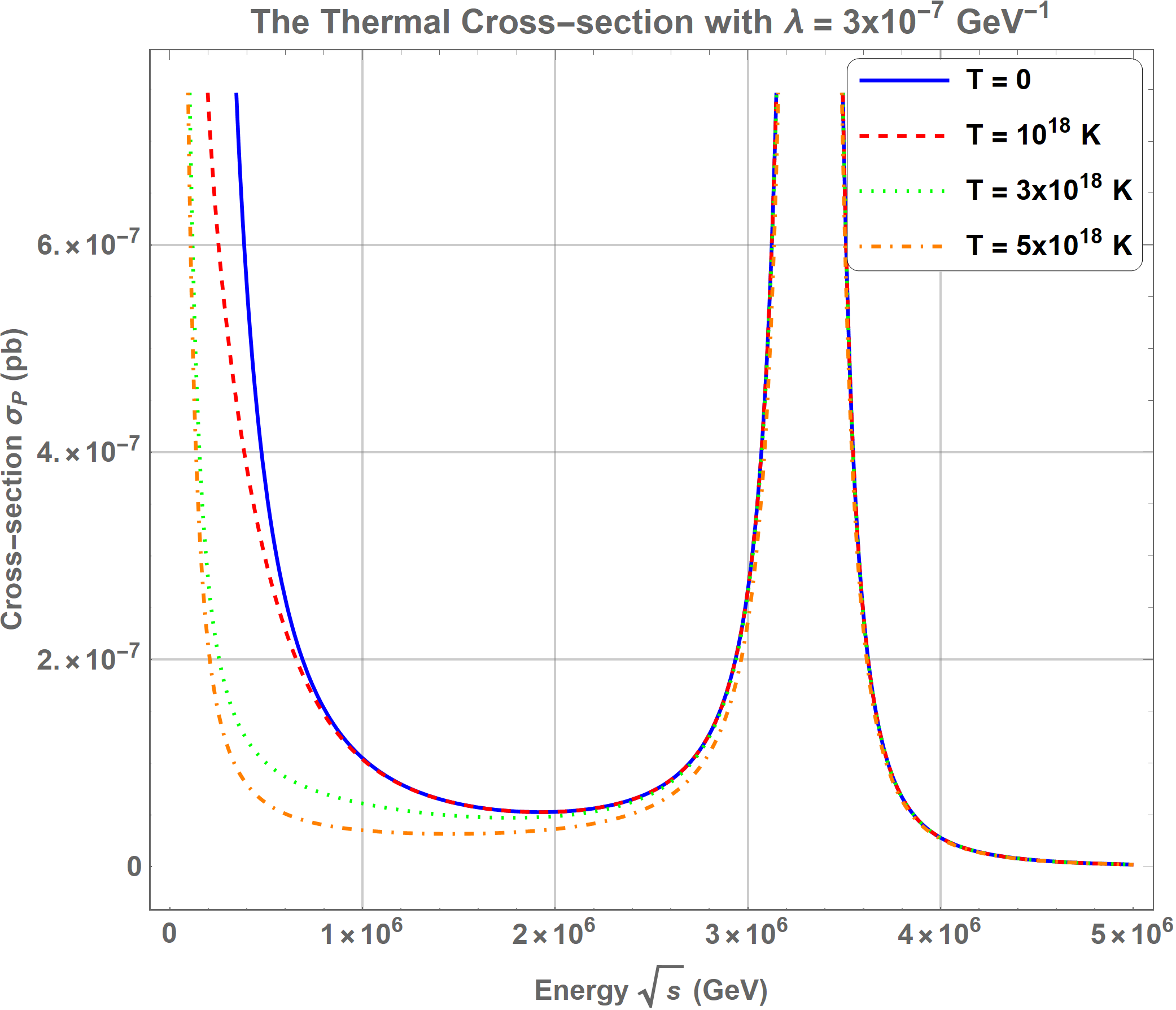}
    \caption{Thermal cross-section \eqref{eq10} versus center-of-mass energy $\sqrt{s}$ at $T=0$ (blue solid line), $10^{12}$ K (red dashed line), $10^{16}$ K (green dotted line), and $10^{18}$ K (orange dot-dashed line). All curves are plotted with $\lambda = 3 \times 10^{-7}$ GeV$^{-1}$.}
    \label{fig3}
\end{figure}

Within the TFD formalism, the differential cross section in the ultrarelativistic limit is written as
\begin{align}
    \left(\frac{d\sigma (\beta)}{d\Omega}\right)_P=\frac{1}{64 \pi^2 s} \langle|\mathcal{M}(\beta)|^2\rangle.
\end{align}
By substituting Eq.~\eqref{MB1} into the expression above, the differential cross section is obtained as
\begin{eqnarray}
     \left(\frac{d\sigma(\beta)}{d\Omega}\right)_P&=&\frac{ e^4(\cos2\theta+3)}{512\pi^2 E^2(1-4\lambda^2E^2)^2}\left(\frac{1+\tanh{\frac{\beta E}{2}}}{2}\right)^4\nonumber\\
     &\times&\left(64\pi^2 \delta^2(s) n(q;\beta) E^4+1\right),
\end{eqnarray}
which can be integrated over the solid angle to yield the total cross section for the interaction
\begin{align}
    \sigma_P(\beta)=\frac{4\alpha^2\pi}{3s(1-4\lambda^2s)^2}\left(\frac{1+\tanh{\frac{\beta \sqrt{s}}{4}}}{2}\right)^4.\label{eq10}
\end{align}
The above expression represents the cross section for the $e^-+e^+\to\ell^-+\ell^+$ scattering process in generalized QED in the presence of thermal effects. Figures \ref{fig2} and \ref{fig3} illustrate this result through various analyses.

Figure \ref{fig2} displays the behavior of the thermal contributions alone, which directly influence the intensity of the scattering process.  According to the form of the thermal function $U^8(\beta)$ given in Eq.\eqref{eq11}, increasing temperature reduces the effective cross-section area of the reaction. This effect becomes noticeable at low energies only when $T > 10^{12}$ K; below this temperature, changes are minimal. In summary, the zero-temperature limit $\beta \to \infty$ recovers the cross section from Eq.\eqref{sigma}, while in the high-temperature limit $\beta \to 0$, the thermal function dominates and significantly modifies the scattering cross section.

Figure \ref{fig3} illustrates the combined effects of the Podolsky parameter $\lambda$ and temperature on the scattering process. As in the zero-temperature case, the presence of $\lambda$ introduces a pole in the cross section, which remains unaffected by temperature. However, temperature plays a significant role in regions away from the pole. As discussed in \cite{pod3}, a lower bound on the photon mass of approximately $37$ GeV is proposed in the context of the primordial universe, corresponding to a pole in Eq.~\eqref{eq10} near this energy scale. This implies a Podolsky parameter value of $\lambda \approx 2.7 \times 10^{-2}$ GeV$^{-1}$, which is larger than the values considered in this work. The values of $\lambda$ shown in Figures \ref{fig1} and \ref{fig3} were chosen arbitrarily as overestimates for illustrative purposes only.

Finally, the motivation for studying temperature effects in a field theory lies in the fact that thermal effects are intrinsic to the environment, and fundamental quantities must exhibit $\beta$-dependence to be properly described. Such thermal characteristics become increasingly important at very high temperatures. The results obtained here indicate that, in the early universe, where particle energies were on the order of $10^{19}$ GeV and temperatures reached approximately $10^{32}$ K, the factor $U^8(\beta)$ reduces the interaction intensity to about 34\% of the value predicted by standard QED with $\lambda=0$.

\section{Conclusions}\label{s4}

Podolsky's electrodynamics has been formulated as a generalization of QED by incorporating a massive sector for the photon while preserving gauge invariance -- unlike Proca's theory. As a result, the high-frequency divergences that arise in standard QED are regularized. In this work, Podolsky's framework was employed to derive the modified photon propagator, which was then applied to the $e^- + e^+ \to \ell^- + \ell^+$ scattering process. Both the probability density and the total cross section were calculated, and, as expected, the resulting cross section was found to depend explicitly on the Podolsky parameter.

The main focus of this work was the investigation of thermal effects on the $e^- + e^+ \to \ell^- + \ell^+$ scattering process. Finite-temperature contributions were introduced through the TFD formalism, in which vacuum expectation values are replaced by statistical averages. To properly implement TFD, the standard Hilbert space was doubled into tilde and non-tilde sectors, and a Bogoliubov transformation was applied to redefine the field operators as linear combinations of thermal operators in both sectors. These constructions allowed the Podolsky photon propagator to be derived at finite temperature, directly modifying the scattering process under thermal conditions. As a result, the total cross section acquired a thermal dependence that becomes significant in certain temperature regimes. These findings highlight the importance of incorporating thermal effects into field-theoretical models, as such effects -- though sometimes negligible -- can become dominant and lead to substantial changes in physical observables under extreme conditions, such as those found in the early universe.

\section*{Acknowledgments}

This work by A. F. S. is partially supported by National Council for Scientific and Technological
Development - CNPq project No. 312406/2023-1. D. S. C., L. A. S. E. and L. H. A. R. F. acknowledge financial support from CAPES.


\global\long\def\link#1#2{\href{http://eudml.org/#1}{#2}}
 \global\long\def\doi#1#2{\href{http://dx.doi.org/#1}{#2}}
 \global\long\def\arXiv#1#2{\href{http://arxiv.org/abs/#1}{arXiv:#1 [#2]}}
 \global\long\def\arXivOld#1{\href{http://arxiv.org/abs/#1}{arXiv:#1}}


\end{document}